%% file: paper.tex
\documentclass{iaus}
\usepackage{graphicx}
\usepackage{natbib}
\usepackage{amsfonts}
\usepackage{amsmath}
\graphicspath{{./fig/}}
\usepackage{bm}

\title[Topological constraints on magnetic field relaxation]
{Topological constraints on magnetic field relaxation}

\author[Simon Candelaresi
\& Axel Brandenburg] 
{Simon Candelaresi,
\& Axel Brandenburg}

\affiliation{NORDITA, KTH Royal Institute of Technology and Stockholm University,
Roslagstullsbacken 23, SE-10691 Stockholm, Sweden \\
and Department of Astronomy, AlbaNova University Center,
Stockholm University, SE-10691 Stockholm, Sweden}

\pubyear{2012}
\volume{294}  
\pagerange{111-222}
\jname{Solar and Astrophysical Dynamos and Magnetic Activity}
\editors{A.\ G.\ Kosovichev, E.\ M.\ de Gouveia Dal Pino and Y.\ Yan, eds.}
\doi{}

\include{newcommands}

\begin{document}

\maketitle

\begin{abstract}

Magnetic field relaxation is determined by both the field's geometry and its
topology.
For relaxation processes, however, it turns out that its topology is a much
more stringent constraint.
As quantifier for the topology we use magnetic helicity and test whether it is
a stronger condition than the linking of field lines.
Further, we search for evidence of other topological invariants, which give
rise to further restrictions in the field's relaxation.
We find that magnetic helicity is the sole determinant in most cases.
Nevertheless, we see evidence for restrictions not captured through magnetic helicity.

\end{abstract}

\firstsection 

\section{Introduction}

Geometry and topology of magnetic field lines fundamentally affect their
dynamics \citep{Woltjer-1958-44-9-PNAS, ArnoldHopf1974, ruzmaikin:331,
Taylor-1974-PrlE, fluxRings10, Yeates_Topology_2010,
Yeates_Topology_2011}.
For instance, strongly tied field lines give rise to strong current sheets
which then facilitate magnetic reconnection under which field lines
brake and connect in a different way.
Reconnection for its part, can give rise to ejections of plasma, which
is of particular interest in the case of our Sun.

While the field's geometry has often been appreciated, its topology
has received less attention.
Loosely speaking, topology determines the field's linkage, while geometry
its configuration in space.
Any two field configurations which are topologically different
cannot be transformed one into
the other without breaking field lines, i.e.\ reconnection.

\section{Magnetic helicity}

Magnetic helicity density is the scalar product of the magnetic vector
potential $\AAA$ and the magnetic field $\BB$, i.e.
\EQ
h = \AAA\cdot\BB.
\EN
Its integral over a closed or periodic system, the total magnetic helicity,
\EQ
H=\int\AAA\cdot\BB \ {\rm d}V,
\EN
is a conserved quantity in ideal MHD and in the limit
of vanishing magnetic resistivity \citep{Woltjer-1958-44-9-PNAS}.

Topologically speaking, $H$ is a quantifier for the mutual linkage of magnetic
flux tubes and their internal twist \citep{MoffattKnottedness1969}.
Twisted and linked fields are severely restricted in their dynamics,
in particular their relaxation.
\cite{ArnoldHopf1974} first quantified this restriction in the realizability condition
\EQ
E(k) = 2|H(k)|/k,
\label{eq: realizability condition}
\EN
with the spectral magnetic energy $E(k)$, the spectral magnetic
helicity $H(k)$, and the wave number $k$.
It gives a lower bound for the magnetic energy in the presence of magnetic
helicity.
In its picturesque interpretation as linking of flux tubes it becomes clear
why magnetic helicity imposes restrictions on the magnetic field decay
given by equation \eqref{eq: realizability condition}.
During relaxation, mutually linked field lines cannot freely evolve without
magnetic reconnection.
As long as reconnection is not aided by strong inflows of magnetic fields
into the reconnection zone, it will not occur fast enough for any
appreciable field change or energy loss.
There exist, however, field topologies of linked magnetic field lines which are
not helical and for which equation \eqref{eq: realizability condition}
has no effect.

\begin{figure}[t!]
\begin{center}
\includegraphics[width=.15\columnwidth]{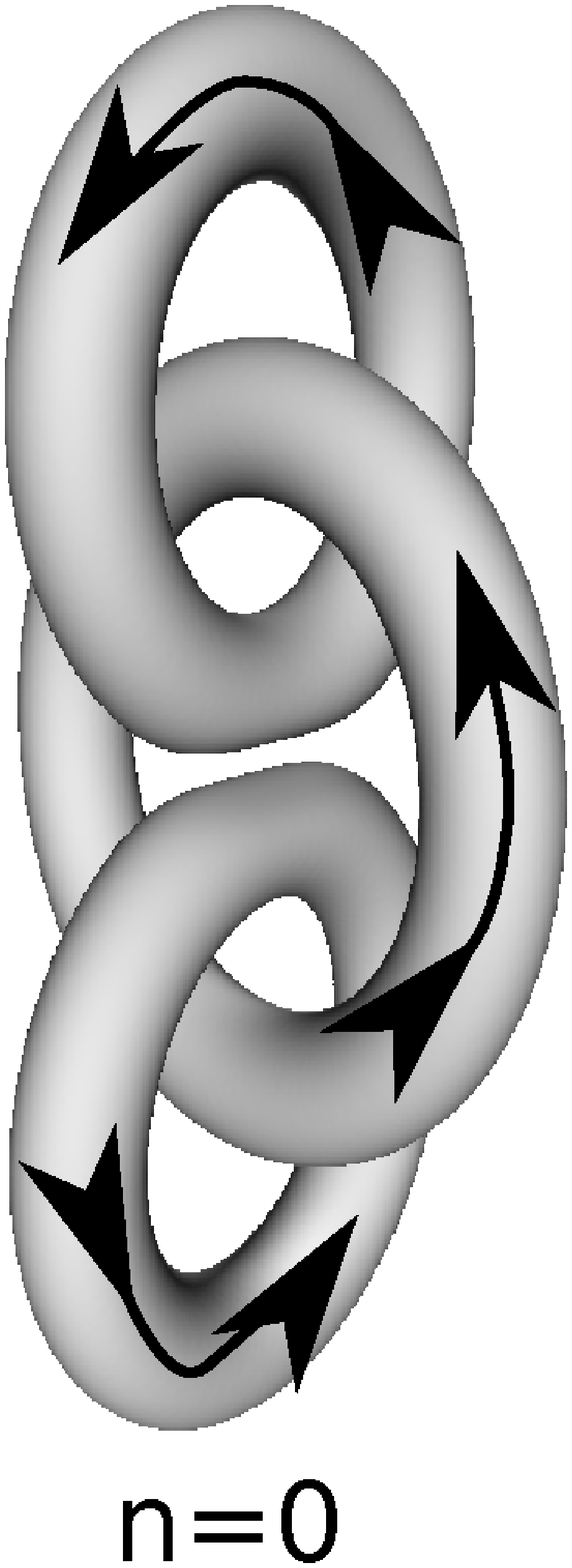} \qquad
\includegraphics[width=.15\columnwidth]{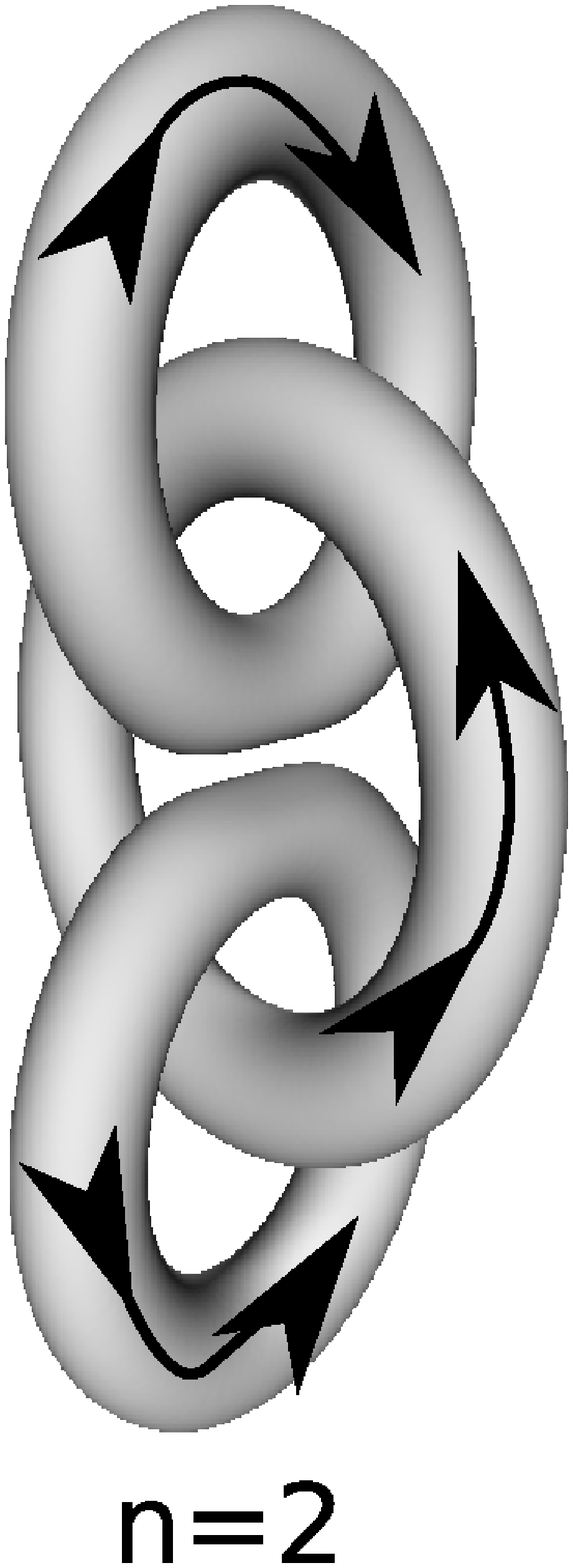} \qquad
\includegraphics[width=.17\columnwidth]{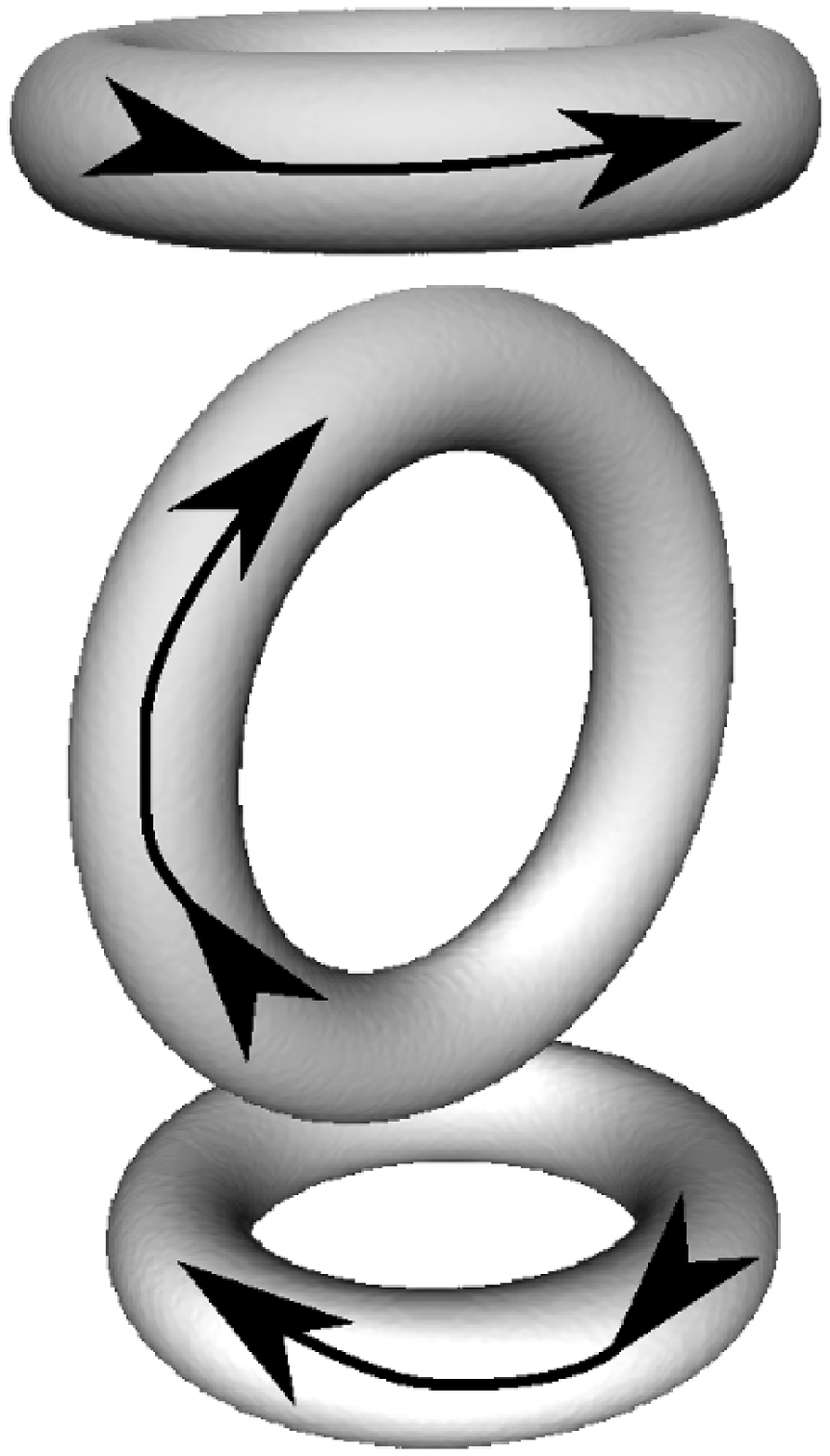}
\end{center}\caption[]{
Iso surfaces of the magnetic energy for the initial magnetic field
configurations.
Arrows denote the direction of the field.
The left configuration is non-helical, while the center one is helical.
The right configuration was used as control run and is not helical.
}
\label{fig: triple rings}
\end{figure}

In a first work we investigate whether the field's topology, as it is given by
the linking and twisting of field lines, is the determining factor in relaxation,
or whether the magnetic
helicity content is the key quantity \citep{fluxRings10}.
From the plethora of possible magnetic field configurations one of the
simplest examples is chosen, which is a triple ring configuration of
interlinked flux tubes (\Fig{fig: triple rings}).

We solve the resistive MHD equations for a viscous, compressible and
isothermal gas
\begin{eqnarray}
&& \frac{\partial}{\partial t} \AAA = \UU\times\BB -\eta\mu_{0}\JJ,
\label{eq: induction} \\
&& \frac{\DD}{\DD t} \UU = -\cs^{2}\nab\ln{\rho} +
\frac{1}{\rho} \JJ\times\BB + \BoldVec{F}_{\rm visc},
\label{eq: momentum} \\
&& \frac{\DD}{\DD t} \ln{\rho} = -\nab\cdot\UU, \label{eq: continuity}
\end{eqnarray}
with the velocity $\UU$,
the molecular magnetic resistivity $\eta$,
the susceptibility in vacuum $\mu_{0}$,
the electric current density $\JJ = \nab\times\BB/\mu_{0}$,
the isothermal speed of sound $\cs$,
the density $\rho$ and the advective time
derivative $\DD/\DD t = \partial/\partial t + \UU\cdot\nab$.
Viscous effects are caught in
$\FF_{\rm visc} = \rho^{-1}\nab\cdot2\nu\rho\SSSS$,
where $\nu$ is the kinematic viscosity,
and $\SSSS$ is the traceless rate of strain tensor with components
${\sf S}_{ij}=\frac{1}{2}(u_{i,j}+u_{j,i})-\frac{1}{3}\delta_{ij}\nab\cdot\UU$.
Commas denote partial derivatives.
Initial magnetic fields represent either of the three configurations in
\Fig{fig: triple rings},
while the initial velocity vanishes in the whole domain and the initial
density is unity.
Boundary conditions are chosen as periodic in order to conserve magnetic
helicity.

\begin{figure}[t!]
\begin{center}
\includegraphics[width=.4\columnwidth]{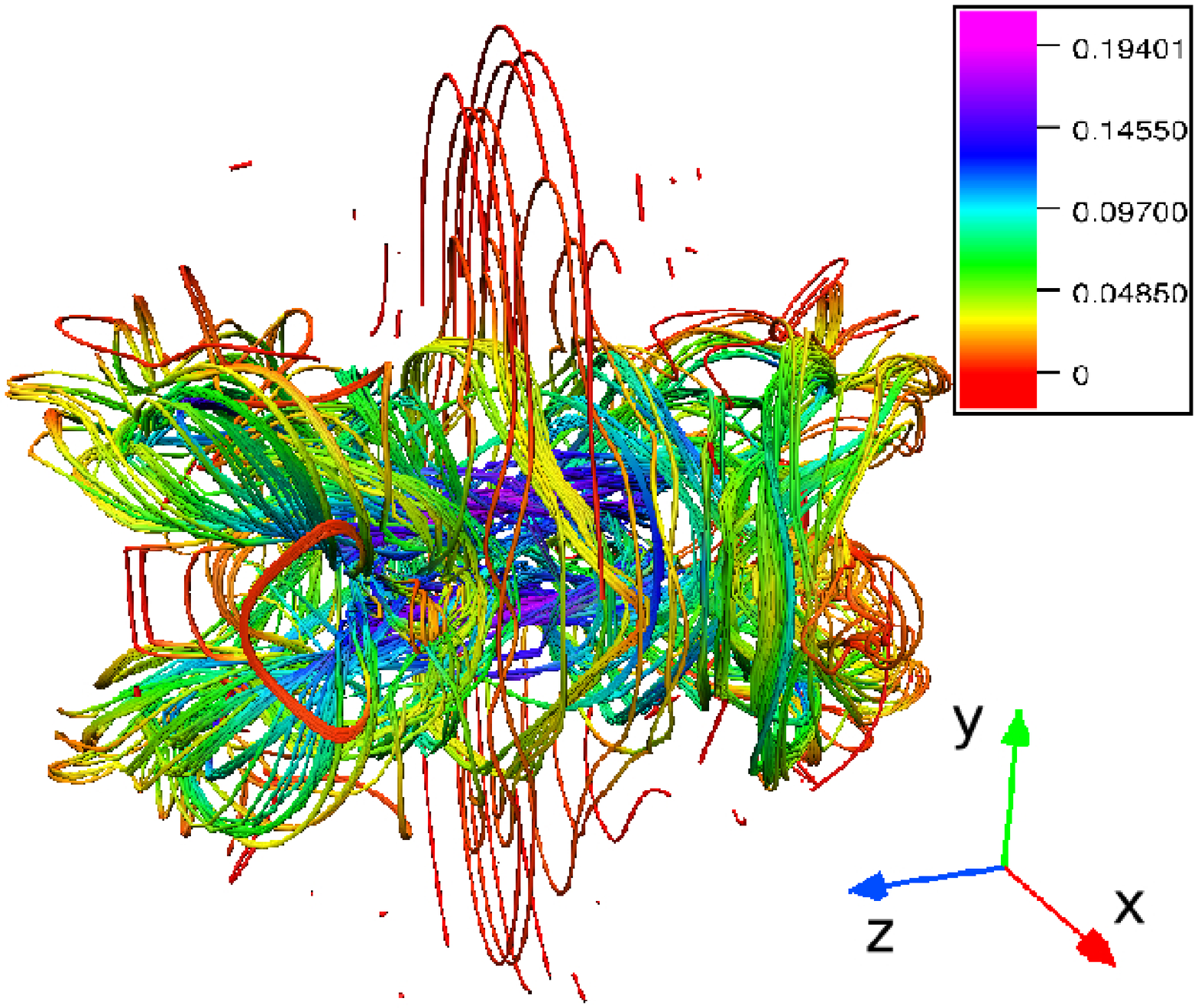}
\includegraphics[width=.4\columnwidth]{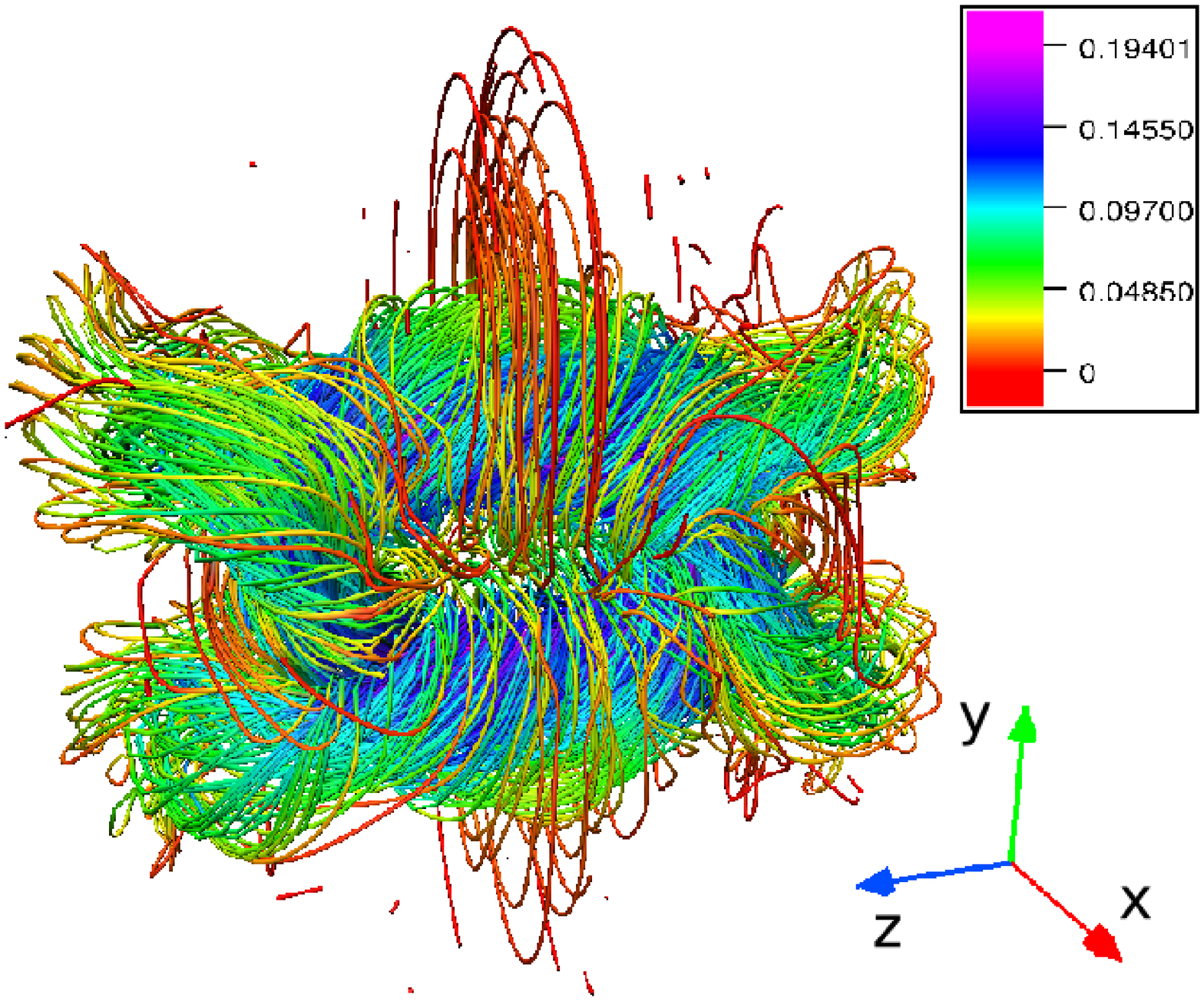}
\end{center}\caption[]{
Magnetic field lines for the two interlocked triple ring configurations
after $4$ Alfv\'enic times of resistive decay.
The non-helical initial configuration (left panel) loses its shape quicker
than the helical configuration (right panel).
}
\label{fig: rings decay}
\end{figure}

\begin{figure}[t!]
\begin{center}
\includegraphics[width=.8\columnwidth]{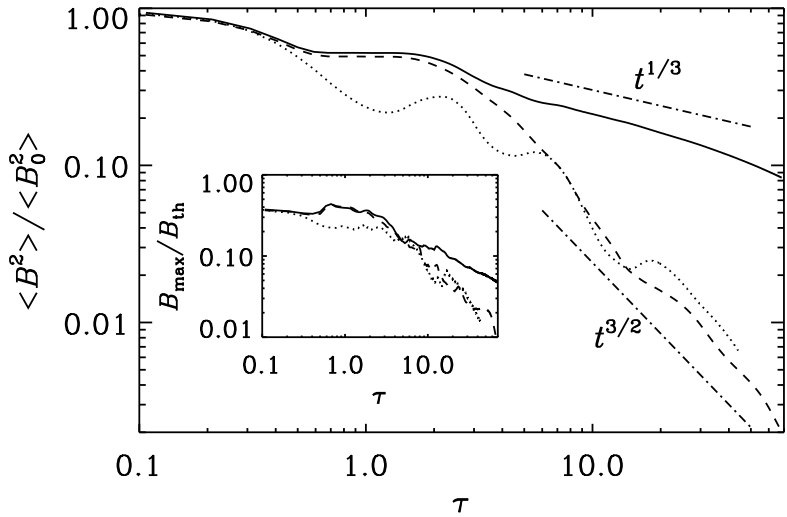}
\end{center}\caption[]{
Normalized magnetic energy evolution for the helical linked rings (solid line),
non-helical linked rings (dashed line) and unlinked rings (dotted line).
}
\label{fig: rings energy}
\end{figure}

From the time evolution of the magnetic field lines it becomes
clear that linking alone cannot hinder the fast decay of the magnetic energy
(\Fig{fig: rings decay}, left panel).
In the presence of magnetic helicity, however, the decay is slowed down
considerably (\Fig{fig: rings decay}, right panel).
Our control setup with non-interlinked flux tubes shows an energy decay
characteristics which is very close to the interlinked non-helical field
(\Fig{fig: rings energy}).
The helical configuration, on the other hand, shows a much slower decay rate.
From this we conclude that magnetic helicity, rather than linking of
flux tubes, determines the field's dynamics.

\begin{figure}[t!]
\begin{center}
\includegraphics[width=.4\columnwidth]{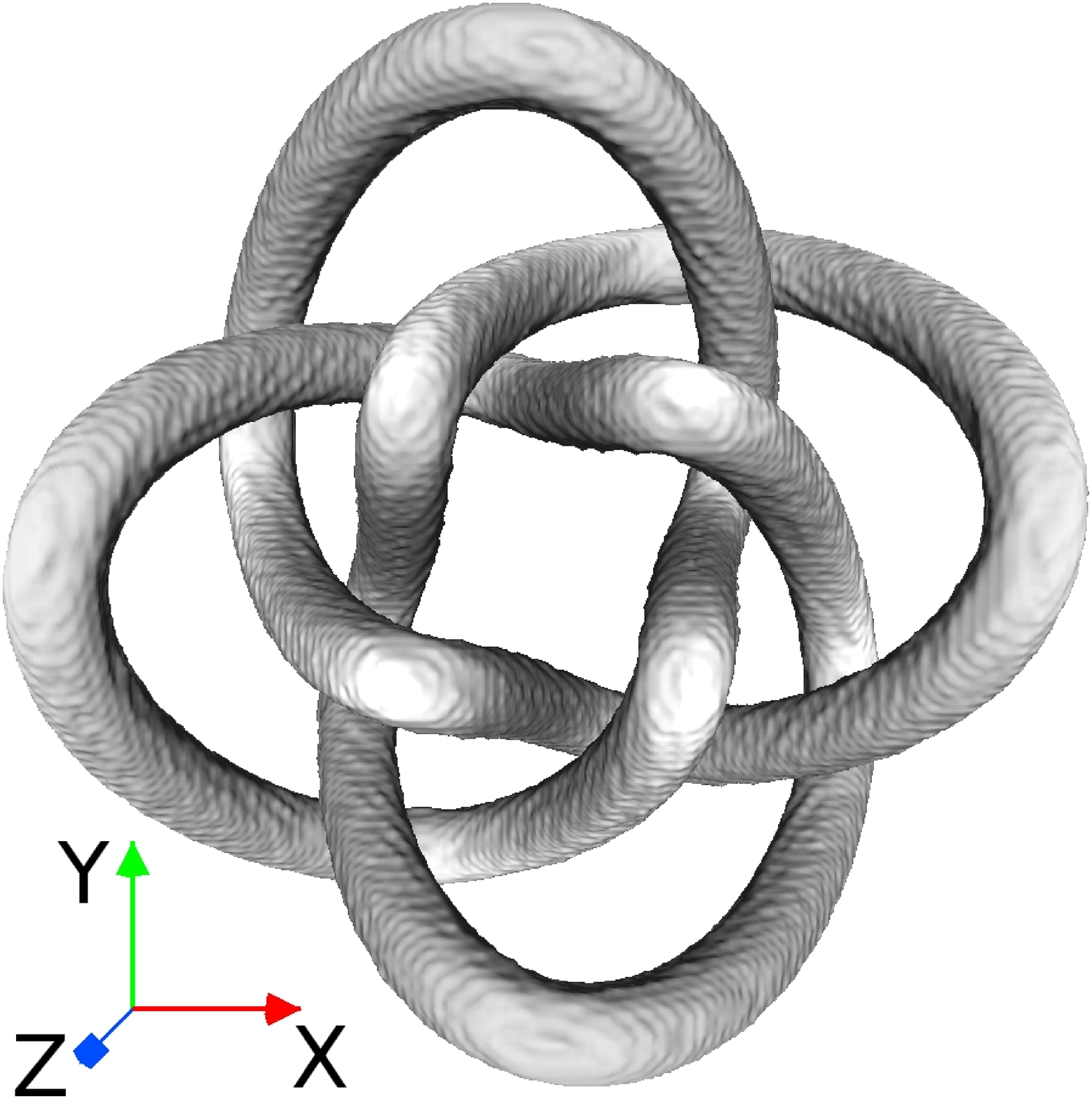} \quad
\includegraphics[width=.4\columnwidth]{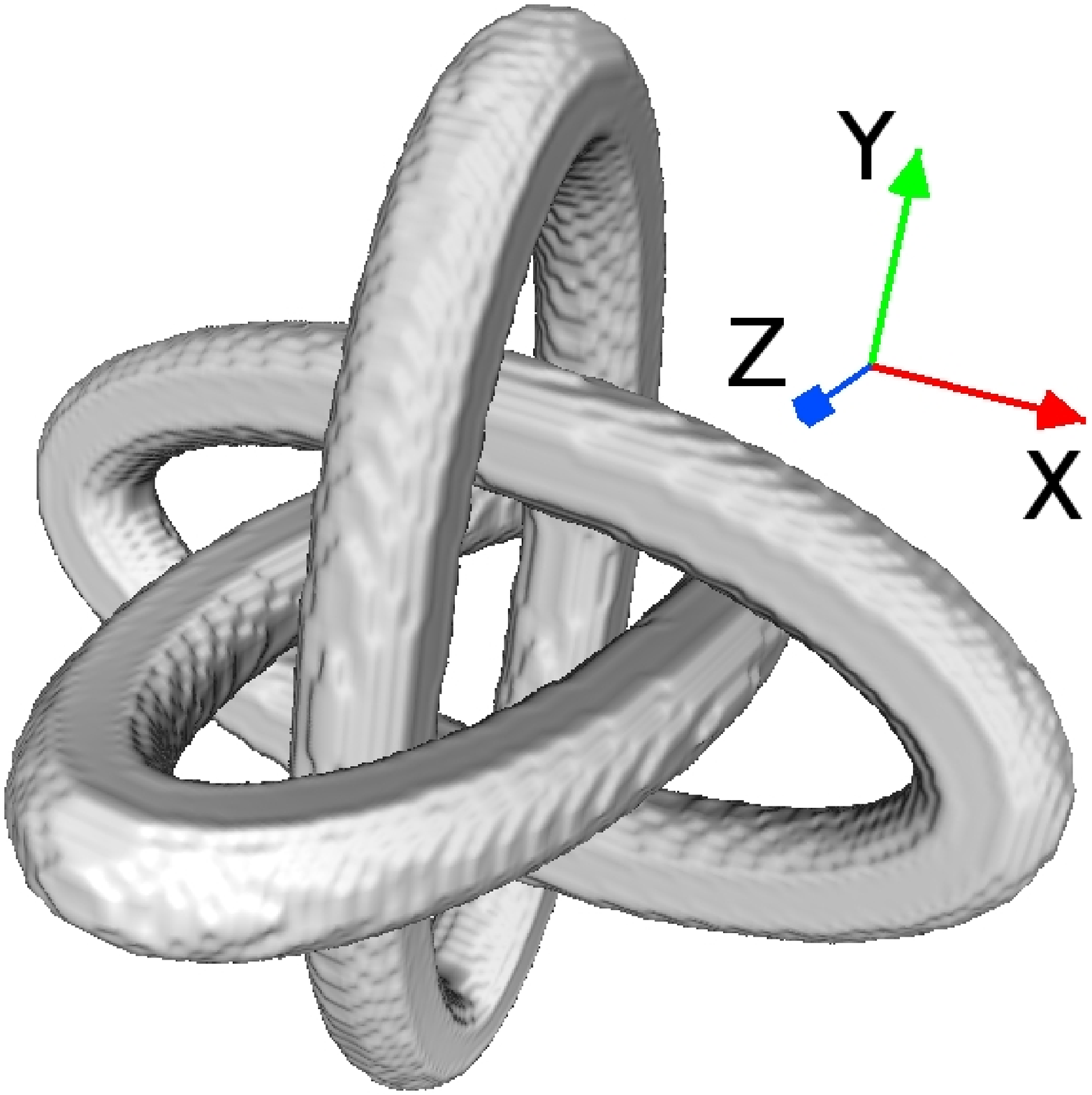}
\end{center}\caption[]{
Iso surfaces of the initial magnetic energy for the IUCAA knot (left panel)
and the Borromean rings (right panel).
}
\label{fig: Borr IUCAA initial}
\end{figure}

\begin{figure}[t!]
\begin{center}
\includegraphics[width=.8\columnwidth]{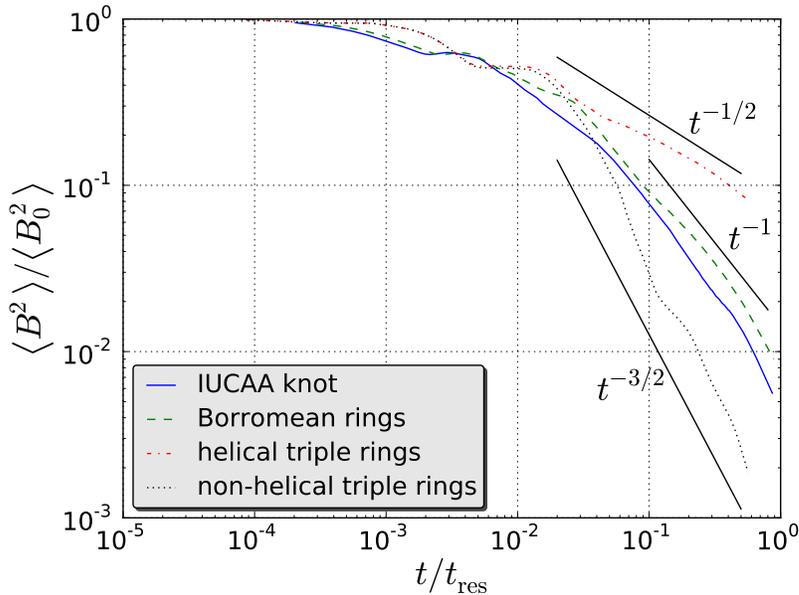}
\end{center}\caption[]{
Comparison of the magnetic energy evolution for different initial field
configurations.
}
\label{fig: knots compare}
\end{figure}

In the same fashion we test the importance of the field's linkage and
knottedness for non-helical configurations with other highly non-trivial topologies.
Initial fields are the IUCAA knot (\Fig{fig: Borr IUCAA initial}, left panel),
which, in the Alexander-Briggs notation, is the
8\_18 knot, and the Borromean rings (\Fig{fig: Borr IUCAA initial}, right panel),
after the emblem of the north Italian aristocratic house of Borromeo.
We compare the magnetic energy evolution with the triple ring configurations
and find that both, the IUCAA knot and the Borromean rings, show an
intermittent power law in the energy decay (\Fig{fig: knots compare}).
This allows for speculations about higher order topological invariants, which
are non-zero for those field configurations and impose additional restrictions on
the field's dynamics; see e.g.\ \cite{ruzmaikin:331}.


\section{Beyond magnetic helicity}

There exists an infinite number of topological invariants for
three-dimensional vector fields.
Applications on MHD have been, nevertheless, very limited and successful
in only a few attempts.
Two invariants of third and forth order in $\BB$ are finite for the Borromean rings
\citep{ruzmaikin:331}.
Their definition is, nevertheless, limited to distinct flux tubes which do
not overlap.
In resistive MHD magnetic field diffuses and an initially confined field
will occupy the whole space.

One way around this hitch is by using the fixed point index
(see e.g.\ \cite{Frankel04, Yeates_Topology_2010}),
which in turn, is only applicable to fields with a preferential direction,
like in toroidal fields, or
a field with a positive $z$-component.
In the latter case one can define a mapping between the bottom and top
boundaries by tracing the field lines, resulting in the field line mapping,
\begin{eqnarray}
 & \Re^{2} \rightarrow \Re^{2}, \\
 & (x,y) \rightarrow \FF(x,y),
\end{eqnarray}
with the initial point $(x,y)$ at $z = 0$.
Note that $\FF(x,y)$ is bijective.

Fixed points are those points for which $\FF(x,y) = (x,y)$.
There can be infinitely or finitely many, with at least one fixed point.
Considering the fixed point's neighborhood in the $xy$ plane we can
determine its sign.
Depending if $\FF^{x}(x,y) > x$ and $\FF^{y}(x,y) > y$, a color is assigned;
for $\FF^{x}(x,y) < x$ and $\FF^{y}(x,y) > y$ a different color is assigned,
likewise for the other cases.
The sequence of these colors around the fixed point determines its sign
$t_{i}$, where $t_{i} \in \{-1,1\}$.
Summing over all fixed points yields the fixed point index
\citep{Brown71, Frankel04}
\EQ
T = \sum_{i} t_{i},
\EN
which is a conserved quantity in ideal MHD \citep{Brown71}.

Simulations using the fixed point index as constraining quantity were performed
by \cite{Yeates_Topology_2010, Yeates_Topology_2011},
where they observed a constraint relaxation of magnetic fields.
Their equilibrium state turned out to be of higher energy than that proposed
by \cite{Taylor-1974-PrlE}.

\section{Conclusions}

From resistive MHD simulations of relaxing interlinked magnetic fields
it becomes apparent that magnetic helicity, rather than actual linkage,
determines the field's relaxation properties.
The decaying IUCAA knot and Borromean rings show some intermittent decline
speed for the magnetic energy.
This suggests that there might be higher order topological invariants,
which impose restrictions on the field's dynamics.
An example of such invariants is the fixed point index,
which is conserved in ideal MHD and is shown to impose further restrictions
on relaxation.

\end{document}

%% file: newcommands.tex
\newcommand{\BoldVec}[1]{\mathchoice%
  {\mbox{\boldmath $\displaystyle     #1$}}%
  {\mbox{\boldmath $\textstyle        #1$}}%
  {\mbox{\boldmath $\scriptstyle      #1$}}%
  {\mbox{\boldmath $\scriptscriptstyle#1$}}%
}
\newcommand{\EQ}{\begin{equation}}
\newcommand{\EN}{\end{equation}}
\newcommand{\EQA}{\begin{eqnarray}}
\newcommand{\ENA}{\end{eqnarray}}

\newcommand{\Fig}[1]{Fig.~\ref{#1}}

\newcommand{\UU}{\BoldVec{U} {}}

\newcommand{\BB}{\BoldVec{B} {}}

\newcommand{\AAA}{\BoldVec{A} {}}

\newcommand{\JJ}{\BoldVec{J} {}}

\newcommand{\FF}{\BoldVec{F} {}}

\newcommand{\nab}{\BoldVec{\nabla} {}}

\newcommand{\SSSS}{\bm{\mathsf{S}}}

\newcommand{\DD}{{\rm D} {}}

\def\cs{c_{\rm s}}